\documentclass[aps]{revtex4}

\input epsf.sty
\newcommand{\insertfig}[2]{\mbox{\epsfxsize=#1cm \epsfbox{#2.eps}}}

\begin{document}

{\flushright \it DOE/ER/40762-268\\
UM PP03-019}

\title{\large \bf  Resolving the Large-$N_c$ Nuclear Potential Puzzle}

\vspace{15mm}

\author{\bf Thomas D. Cohen}

\affiliation{ Department of Physics,
 University of Maryland,\\
 College Park, MD 20742-4111,USA \vspace{.15in}}

\begin{abstract}The large $N_c$ nuclear potential puzzle arose
because three- and higher-meson exchange contributions to
the nucleon-nucleon potential did not automatically yield
cancellations that make these contributions consistent with
the general large $N_c$ scaling rules for the potential.
Here it is proposed that the resolution to this puzzle is
that the scaling rules only apply for energy-independent
potentials while all of the cases with apparent inconsistencies
were for energy-dependent potentials.  It is shown explicitly
how energy-dependent potentials can have radically different
large $N_c$ behavior than an equivalent energy-independent one.
One class of three-meson graphs is computed in which the contribution
to the energy-independent potential is consistent with the
 general large $N_c$ rules even though the energy-dependent
potential is not. \vspace{.25in}
\end{abstract}

\maketitle

\section{Introduction}
 The nature of the nucleon-nucleon
interaction is at the heart of nuclear physics.  One traditional
picture of nuclear interactions at low energies is that they are
mediated via meson-exchange. From the QCD perspective one can
envision the quarks and gluons organizing themselves into hadrons
and then the baryons interact amongst themselves via the exchange
of mesons. It is not immediately clear how one can test this
picture how nucleon-nucleon interactions emerge from QCD since we
have no {\it a priori} method for deriving these  interactions
directly from QCD.  In ref.~\cite{BanCohGel01} it was pointed out
that large $N_c$ QCD can provide some insight into the issue.
Since the meson-exchange picture of the nucleon-nucleon
interaction if valid is justified on rather generic grounds, it
should be expected to be hold at any $N_c$ greater than unity and
hence should hold for large $N_c$.  However, as noted in ref.
\cite{BanCohGel01} it is by no means obvious that the
meson-exchange picture is in fact consistent with large $N_c$
counting rules.  In particular, there is a threat that while a
one meson-exchange description yields nucleon-nucleon
interactions which are consistent with the large $N_c$ counting
rules\cite{KapMan96,KapSav95},  multiple-meson exchange graphs
yield interactions which are not.  A consistent large $N_c$
description requires a cancellation of all of these dangerous
graphs.  It was shown in ref.~\cite{BanCohGel01} that for all
two-meson exchange such cancellations do in fact occur, provided
that the large $N_c$ scaling rules for the nucleon-nucleon
interaction are interpreted as applying for a nucleon-nucleon
potential for use in a Sch\"odinger or Lippmann-Schwinger
equation (as opposed to a kernel in a four dimensional
Bethe-Salpeter type equation). This result seems to support the
traditional meson-exchange picture for nucleon-nucleon
potentials. Clearly, this support would be stronger if the
cancellations seen for two-meson exchange also happen for general
multiple-meson exchanges.

Unfortunately, in ref.~\cite{BelCoh02} it was shown that the extension of the techniques of
ref.\cite{BanCohGel01} does not automatically lead to the types of cancellations seen for
two-meson exchanges.  Thus, in the absence of some type of conspiracy leading to such
cancellations there appears to be no way to justify the meson-exchange picture from large
$N_c$ QCD.  The result is puzzling---why should such cancellations occur for two-meson
exchange for multiple channels and at the same time fail for three-or-higher meson exchange?
The purpose of this note is to resolve this puzzle. As will be argued below, there is a
reorganization of the analysis that yields precisely the type of conspiracy needed to
resolve the puzzle.

To begin with let us consider the background to the problem. Large $N_c$
QCD\cite{Hoo74,Wit79}  has proven to be a powerful tool to learn about qualitative and
semiquantitative  features of hadronic physics.  In principle one may hope that it will also
provide important insights into nuclear physics. The possible implications of large $N_c$
QCD for nuclear interactions were already evident in Witten's original paper on baryons in
large $N_c$ QCD.  Witten pointed out that (i) the nucleon-nucleon interaction is
characteristically of order $N_c$, (ii) that nucleon-nucleon scattering observables had no
smooth limit as $N_c \rightarrow \infty$ if the nucleon momenta were of order $N_c^0$, and
(iii) for momenta of order $N_c$, a relativistic time-dependent Hartree formalism is
appropriate and has a smooth large $N_c$ limit.  In practice, no such time-dependent Hartree
calculations have been carried out, although recently\cite{CohGel02} it was shown what type
of observables are calculable in principle in this framework
and the spin- and isospin-dependence of these observables was deduced.

The problem of nucleon-nucleon interactions for momenta of order $N_c^0$ is of real
significance.  Kaplan and Manohar\cite{KapMan96} following the work of Kaplan and
Savage\cite{KapSav95} have argued that useful information about the nucleon-nucleon
potential can be extracted in this regime.  In particular, they
suggest that the nucleon-nucleon
potential can be associated with quark-line connected diagrams between two color-singlet
clusters of $N_c$ quark lines. This is motivated by Witten's Hartree analysis.  Combining
this with the known contracted SU(4) spin-flavor symmetry of two flavor
QCD\cite{GerSak83,GerSak84,DasMan93,DasJenMan93,Jen93} for the coupling to each cluster, this was used to
deduce that the large $N_c$ scaling behavior for the various spin and isospin contributions
to the potential.  The leading scaling behavior is given by
\begin{equation}
V_{I=J} \sim
N_c \; \; ; \; \;V_{I \ne J} \sim N_c^{-1} \; , \label{KSM}
 \end{equation}
where the subscript indicates the quantum numbers of the exchange in the $t$ channel.  It is
worth noting that nucleon-nucleon potentials which are fit to scattering data have a pattern which is
consistent with eq.~(\ref{KSM}) in the sense that the components which are order $N_c$ in
eq.~(\ref{KSM}) are characteristically significantly larger than those which are of order
$1/N_c$ \cite{KapMan96,Ris02}. Recently there have also been attempts to study this regime
from the perspective of effective field theory \cite{Bea02}

The issue of consistency is studied in the following way.  First one supposes that there exists
some hadronic field theory whose masses and couplings
scale with $N_c$ according to the standard
large $N_c$ rules\cite{Coh89}.  It was
pointed out in refs.\cite{BanCohGel01,Ris02} that the large
$N_c$ scaling rules of meson-baryon couplings given by the contracted SU(4) spin-flavor symmetry
of two flavor QCD\cite{GerSak83,GerSak84,DasMan93}  yield a
one-meson exchange potential and will
satisfy eq.~(\ref{KSM}).  At two-meson exchange the key issue is that both the box-graph and the
crossed-box graph are formally of order $N_c^2$ and hence individually cannot be consistent with
eq.~(\ref{KSM}).  It is straightforward to show, however that the box graph can be decomposed into
two parts---a contribution arising from the nucleon poles and a contribution arising from the
meson poles when the graph is evaluated via contour integration.  The contribution arising from
the nucleon poles can be shown to be of exactly the same form as an iterate of the one-meson
exchange potential in a Lippmann-Schwinger equation.  Thus, the nucleon pole contribution to the
graph will be picked up when solving the Schr\"odinger or Lippmann-Schwinger equation and must not
be included as part of the potential  in order to avoid double counting.  The meson-pole
contributions to the box graph are retardation effects.  The full contribution to the potential
from these graphs is the sum of
the retardation part of the box-graph with the crossed-box.

In ref.~\cite{BanCohGel01}, it was shown that for all types of two-meson exchange there are
cancellations between these two graphs so that the sum is consistent with eq.~(\ref{KSM}).  These
cancellations were highly nontrivial since a number of different spin and isospin structures must
all cancel.  For example,
two-pion exchange, for which both the box and crossed-box
contributions are of order $N_c^2$, contribute to,
among other things, the isoscalar central potential, which is of
order $N_c$ and requires cancellations of at least relative
order $1/N_c$ and to the isovector
central potential, which is of order $1/N_c$ and requires cancellations of a least relative order
$1/N_c^3$, in order to maintain consistency.  By explicitly checking meson exchanges for all
relevant spin and isospin couplings it was seen that all of the ``dangerous'' contributions that
were inconsistent with eq.~(\ref{KSM}) canceled to the degree necessary to ensure consistency.
This demonstration required explicit use of the contracted SU(4)
algebra  which in turn implied
that intermediate $\Delta$ states have to be kept as explicit degrees of freedom in the potential
model to obtain consistency.

The analysis of ref.~\cite{BanCohGel01} clearly helps justify the meson exchange picture of
nucleon-nucleon forces.  However, no general theorem was proved.  Rather all of the relevant cases
for two-meson exchange were individually tested.  Clearly, one's confidence in the generality of
the result would increase if a number of examples of three- and higher-meson exchange show the
same type of cancellations seen in two-meson exchange.  An analysis of certain multi-meson
exchange graphs was done in ref.~\cite{BelCoh02}.  The diagrams involved can get quite
complicated.  Accordingly, it was necessary to establish some bookkeeping rules.  The basic method
used was in many ways analogous to that used in ref. ~\cite{BanCohGel01}: First one identifies all
two-baryon irreducible Feynman diagrams as contributing to the potential.  Next one considers the
two baryon-reducible graphs and notes that such graphs all contain parts that have two baryon
propagators between interactions.  These propagators
are then expressed  as the sum of two
parts---one where one of the nucleons is on-shell,
and the remainder.  Next, one makes use of the
fact the two baryon propagators with one baryon on-shell is identical to the propagator in the
context of a Lippmann-Schwinger equation (up to relativistic corrections that are suppressed by
$1/N_c$).   Thus, these contributions will be included when iterating the Lippmann-Schwinger
equation the potential is identified as
coming from the two-baryon reducible graphs as the full
graph minus the contributions arising from two propagators between interactions with one baryon on
shell.

This organization of the full problem into a potential and then its iteration via
a Lippmann-Schwinger or Schr\"odinger equation has a number of virtues.  First,
provided the problem is nonrelativistic, the scattering amplitude obtained by
such a procedure correctly reproduces the sum of all Feynman diagrams for on-shell
scattering.  Secondly this procedure is well suited to the study of ladder and
crossed-ladder graphs including various non-commuting couplings since the non-abelian
generalization of the eikonal
formula of refs.~\cite{LamLiu96,LamLiu97} can be
implemented for the sum of these graphs.  The non-abelian generalization of the
eikonal formula expresses the sum of all meson and crossed-meson lines entering
a single baryon in terms of commutators multiplying $\delta$ function which
correspond to an on-shell baryon.  In the context of summing ladder and crossed-ladder
contributions to baryon-baryon scattering it is straightforward
to see which of these on-shell
contributions correspond to iterates of the Lippmann-Schwinger equation.

\begin{figure}[t]
\begin{center}
\mbox{
\begin{picture}(0,140)(160,0)
\put(0,0){\insertfig{11}{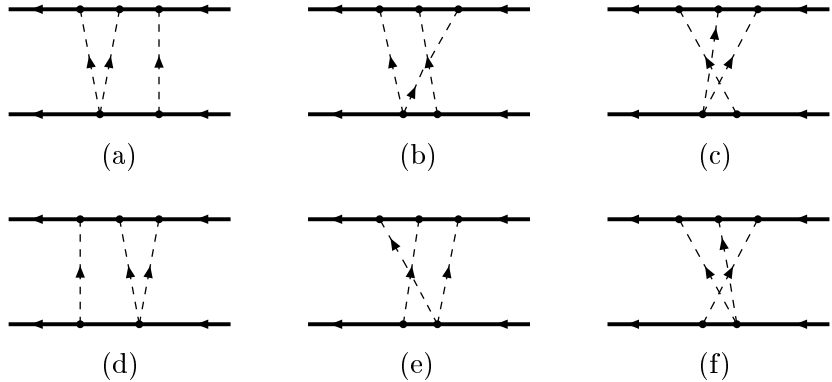}}
\end{picture}
}
\end{center}
\caption{\label{seagull}Three scalar-isoscalar meson exchange diagrams with one seagull type coupling.}
\end{figure}

When this organizing principle was implemented for multi-meson
exchanges, however, it was found that the cancellations needed
for the potentials to maintain consistency with eq.~(\ref{KSM})
did not occur.  This was seen for two classes of graphs that were
considered.  One class was the sum of ladders and crossed-ladders
with non-commutating couplings. The most dangerous type was where
the meson coupled to baryon in a vector-isovector manner. The
issue there was the emergence of contributions in which one
nucleon was on-shell but which were not iterates of the potential
as defined above.  The generalized eikonial
formula\cite{LamLiu96,LamLiu97} implies that these yield
contributions to the potential  given in terms of commutators.
Although the commutators typically yield a $1/N_c^2$ suppression
and may be expected to induce a commutator on the other nucleon
leg for another factor of $1/N_c^2$ if the ladder has six rungs
or higher, the explicit $N_c^{1/2}$ associated with each
meson-nucleon vertex overwhelms the suppression and an
inconsistency with eq.~(\ref{KSM}) is the result.  A simpler case
where an inconsistency can be seen is the case of three
scalar-isoscalar meson exchange between nucleons where two of the
mesons couple to one of the nucleons in a seagull type vertex as
shown in fig.~(\ref{seagull}).  Again when the part representing
an iterate of the potential using the organizing principle
discussed above is removed, the remaining contribution which
contributes to the potential itself does not cancel and is of
order $N_c^2$, in violation of eq.~(\ref{KSM}).  This
inconsistency is the large $N_c$ potential puzzle.

In the remainder of this article a resolution to the puzzle will
discussed.  In the following section, it will be suggested that
the heart of the problem lies in the organizing principle
discussed above which has the feature that the potential obtained
have explicit energy dependence. Following this a toy problem
will presented to show how energy dependence can alter the the
$N_c$ counting of a potential.  Next a new organizing principle is
suggested which yields energy independent potentials. In the
final section it is shown explicitly that this new procedure, when
applied to the case of three scalar-isoscalar meson exchange
where two of the mesons couple to one of the nucleons in a
seagull type vertex (of the type in fig.~(\ref{seagull})), yields
a potential consistent with eq.~(\ref{KSM}).

\section{Energy dependence}
 In ref. \cite{BelCoh02} various possible
resolutions to the puzzle were suggested. One possibility was
that necessary cancellations might happen naturally and
generically (without any special conspiracies involving coupling
constants of different mesons) with a different organization of
the problem. However, no plausible reorganization was suggested.
The purpose of the present paper is to argue  that this is the
correct resolution of the
 problem and to provide the needed reorganization.  It will then be shown explicitly that
with this reorganization,
the contributions to the potential of the three-meson exchange graphs of fig.~(\ref{seagull})
indeed give rise to the cancellations needed for the consistency with eq.~(\ref{KSM}).

The key to this reorganization is the realization that the separation of the contributions
into a potential and its iterates in a Lippmann-Schwinger equation is not unique if one allows
energy-dependent potentials: there are an infinite number of ways to distribute the energy
dependence between explicit energy dependence of the potential and energy dependence arising
from iteration which yield identical scattering amplitudes.  Of course, one often thinks of
potentials as being energy independent.  However, energy-dependent potentials arise naturally
when one suppresses explicit inclusion of degrees of freedom and in the present problem the
final NN potential suppresses explicit meson degrees of freedom that are in the underlying problem.
The procedure used in ref.~\cite{BelCoh02} described above to isolate the potential from its
iterates produces such an energy-dependent potential. Of course, at a fundamental level, there
is nothing wrong with energy dependent potentials for use in a Schr\"odinger equation provided
they correctly predict the physical scattering amplitudes, and the procedure used in ref.~\cite{BelCoh02}
should reproduce the scattering amplitude.  The notion that the elimination of energy dependence to establish
consistency may remind one of the situation in NRQCD where field redifintions are used.\cite{ManStew}

Since the correct physical amplitude is obtained, the fact that the
procedure used in ref.~(\cite{BelCoh02})
produces an energy-dependent potential might seem completely innocuous.  However, the key point is that
the large $N_c$ scaling rules in eq.~(\ref{KSM}) is {\it not} for a physical quantity which can be
directly measured; rather it is for the potential.
Moreover, it is possible for an energy-dependent
potential to have radically different $N_c$ scaling behavior from an energy-independent potential
which is physically equivalent.

\section{A toy problem}
 To see how  energy dependence can alter the $N_c$ scaling consider
the following simple example: Begin with a simple
energy-dependent potential operator,
 \begin{equation}
\tilde{V}(E) = \tilde{V}_0 + \tilde{V}_1 (E - \hat{p}^2/M_N)
\label{example} \, , \end{equation}
 where the tilde is used to distinguish the
potential from an equivalent energy-independent one;
$\tilde{V}_0$ and $\tilde{V}_1$ have no energy dependence (so
that all energy dependence is explicitly given), $\hat{p}$ denotes
the relative momentum operator, and $M_N/2$ is the reduced mass.
The potential is not explicitly Hermitian but this is not
significant for the present purpose, which is merely
illustrative.  This nonhermitian form is used since it is the
simplest model which demonstrates the key point. The scattering
amplitude $T$ is obtained via the Lippmann-Schwinger equation,
which as an operator equation can be represented as
\begin{eqnarray}
 T & = & V + V G_{LS} T = V + V G_{LS} V + V G_{LS} V G_{LS} V
+ \ldots \nonumber \\ &{\rm  with}& \; \; G_{LS} = (E - \hat{p}^2/M_N +
i\epsilon)^{-1} \, . \label{LS}
\end{eqnarray}
Now suppose that we have another
potential $V$ which is energy independent and  has the same on-shell scattering
amplitude as in eq.~(\ref{example}).  By comparing the iterates of
eq.~(\ref{LS}) for the two potentials, it is straightforward to see that
\begin{equation}
 V =  \tilde{V}_0 +  \tilde{V}_1 \tilde{V}_0 +
  \tilde{V}_1 \tilde{V}_1 \tilde{V}_0 + \dots
\sum_{ j = 0, \infty}  \tilde{V}_1^j \tilde{V}_0 \, .
\label{equiv}
\end{equation}
The products of the $\tilde{V}_1$ and $\tilde{V}_0$ emerge in $V$
from iterates of the Lippmann-Schwinger equation for $\tilde{V}$;
the $(E - \hat{p}^2/M_N)$ factors accompanying $\tilde{V}_1$ when
multiplied by $G_{LS}$ give unity and thus correspond to an
uniterated term in the Lippmann-Schwinger equation for $V$. Also,
the $(E - \hat{p}^2/M_N)$ operators annihilates the on-shell
external state and thereby eliminates many terms in the sum. Thus,
we see explicitly that energy dependence reshuffles what goes into
the potential and what is obtained via iteration.   For the
present context this is significant in terms of the $N_c$
behavior.  From the form of eq.~(\ref{equiv}), it is easy to see
that the energy-dependent potential $\tilde{V}$ and the
energy-independent potential $V$ cannot both generically be of
order $N_c$.  If $\tilde{V}$ is generically of order $N_c$ ({\it
i.e.} both $\tilde{V}_0$ and $\tilde{V}_1$ are order $N_c$), then
we see that $V$ contains all powers of $N_c$. Conversely, if $V$
and $\tilde{V}_1$ are each of order $N_c$, then $\tilde{V}_0$
contains all powers of $N_c$.

\section{A new organizing principle}
We see from the preceding example that energy-dependent
potentials need not have the same $N_c$ dependence as
energy-independent ones.  The large $N_c$ scaling rules of
eq.~(\ref{KSM}) were derived by associating the quark-line
connected diagrams with ``the potential''.  This association is
somewhat heuristic and it is not immediately obvious whether it is
supposed to apply for energy-dependent or energy-independent
potentials.  It is a reasonable hypothesis, however, that it
applies to energy-independent potentials. Thus, it is plausible
that the failure of the analysis of ref.~\cite{BelCoh02} to
reproduce the $N_c$ scaling rules for eq.~(\ref{KSM}) is because
the separation into a potential and its iterates used in the
analysis does not ensure that the potential is energy
independent.  To see why the potential so derived can depend on
energy, consider the algorithm used in the analysis to isolate
the potentials.  One calculates Feynman diagrams and subtracts
off all contributions coming from places where one nucleon out of
a pair that separates interacting subdiagrams is on-shell.  This
removes terms that look like Lippmann-Schwinger equation
iterates.  The issue is simply that the remaining subdiagrams may,
themselves, be energy dependent.

To test whether energy-independent potentials from multiple meson exchange are consistent with the
rules of eq.~(\ref{KSM}), we must first extract the contributions from energy-independent
potentials from various meson exchanges graphs.  The basic algorithm is similar to the one used in
ref.~\cite{BelCoh02} with one modification.  The potential
contributions from two-particle
irreducible diagrams will be taken as the static limit ({\it i.e.} zero energy limit)  of the
diagram.  For two-particle reducible graphs, one subtracts off all contributions coming from
places where one nucleon out of a pair that separates
interacting subdiagrams is on-shell, and
where the remaining subdiagrams are replaced by their static limits.

It is worth noting at this point that at the level of two-meson
exchange, the organizing principle of ref. \cite{BelCoh02} and
the present one both give results consistent with eq.~(\ref{KSM}).
The reason for this is quite simple.  The two only differ in the
treatment of the energy dependence of subdiagrams.  In the case
of two-meson exchange, the subdiagrams are single meson exchanges
which do have nontrivial energy dependence due to the poles in
the meson propagators.  However, when the external lines are
on-shell, the two methods are identical; both methods have the
same cancellation between retardation effects in the box diagram
and the full crossed-box.  Of course, one can distinguish between
the two-meson exchange in the two methods if one or the other
external line goes off-shell.  However,  external line can go
off-shell only inside larger graphs and the lowest order one can
distinguish between the two approaches is at the level of
three-meson exchange.

\section{Testing the $N_c$ scaling of energy-independent potentials}

The modification suggested in the previous section greatly
complicates the analysis. For involved cases, such as ladders and
crossed-ladders with many rungs, it may require a considerable
effort to test the consistency of eq.~(\ref{KSM}) since the
nonabelian generalization of the eikonal formula can no longer be
implemented in a straightforward way.  However, the case of three
scalar-isoscalar meson exchange between nucleons where two of the
mesons couple to one of the nucleons in a seagull type vertex
represented in fig.~(\ref{seagull}) is tractable.  The diagrams in
fig.~(\ref{seagull}) include all contributions which has the
seagull attached to one of the nucleon lines.  There is an
identical contribution which have the seagull attached to the
other line. It is useful to group the diagrams into two sets,
(a), (b) and (c) as one group and (e), (f) and (g) as the other.
The cancellations needed to get consistency with eq.~(\ref{KSM})
can be shown to occur separately in these groups.  Now let us
consider the contribution from the first three diagrams.  The
amplitude is given by
\begin{eqnarray}
i {\cal A}_{a,b,c} \,&=&\,  g_{1m}^4 g_{2m}
\int \prod_{j = 1}^{3} \frac{d^3 \mbox{\boldmath$k$}_j}{(2 \pi)^3}
\, (2 \pi)^3 \, \delta^{(3)}
\left( \sum_{j = 1}^{3} \mbox{\boldmath$k$}_j
- \mbox{\boldmath$q$} \right)
\int \prod_{j = 1}^{3} \frac{d \omega_j}{2 \pi} \,
(2 \pi) \, \delta \left( \sum_{j = 1}^{3} \omega_j - q_0 \right)
\left( \prod_{j = 1}^3 D (k_j^2) \right) \nonumber \\
& \times & \, G(p - k_1) \, [G(\tilde{p} + k_1 ) G(\tilde{p} + k_1 + k_2) \,
+ \, G(\tilde{p} + k_2) G(\tilde{p} + k_1 + k_2) \, \nonumber \\
&+&  G(\tilde{p} + k_2) G(\tilde{p} + k_2 + k_3)] \, .
\end{eqnarray}
where boldfaced indicates three vectors, and non-boldfaced four
vectors, with $k_j = (\omega_j, \mbox{\boldmath$k$}_j)$;
$g_{1m}$ and $g_{2m}$ are the one-meson and two-meson coupling
constants; the initial four-momenta of the two nucleons are
$p$ and $\tilde{p}$; the four-momentum transfer is
$q$; $D(k) \equiv (k^2 - m_m^2 + i \epsilon)^{-1}$ is the
meson propagator; and the fermion
propagator is denoted by $G$.  For simplicity we will work in the center of mass frame with $p = (
\mbox{\boldmath$p$}^2/ (2 M_n), \mbox{\boldmath$p$} )$  and $\tilde{p}=( \mbox{\boldmath$p$}^2 /(2
M_n), -\mbox{\boldmath$p$} )$.  In fact, the graph as written is ultraviolet divergent.  We will
assume that it is regulated by some short distance physics
which acts to cut off the momentum
integrals.  The details of how this is done is irrelevant for what follows, provided the same
cutoff procedure is used for all of the graphs.

Let us note the $N_c$ dependence of the various inputs to this expression: $g_{1m} \sim
N_c^{1/2}$, $g_{2 m} \sim N_c^0$ and $M_N \sim N_c$; the meson mass in the propagators $D$ is of
order $N_c^0$.  Thus, there is an overall prefactor of $N_c^2$
coming from the coupling constants;
it is this factor which must somehow be canceled up to
relative order $N_c^{-1}$ in order to
get consistency with eq.~(\ref{KSM}).  The kinematic
regime of interest is intrinsically
nonrelativistic since $\mbox{\boldmath$p$}\sim N_c^0$
while $M_N \sim  N_c^1$.  In this regime
it is legitimate to replace the full fermion propagator by
\begin{equation}
G(p) = \frac{1}{p_0   - \mbox{\boldmath$p$}^2 /(2 M_N) + i \epsilon}
\frac{1 + \gamma_0}{2} \, \left ( 1 + {\cal O}(1/N_c) \right ) \, ,
\label{prop}
\end{equation}
where the $1/N_c$ corrections come from the nonrelativistic reduction.
The propagator contains a recoil correction.  This will be a $1/N_c$
correction everywhere except in the vicinity of the propagator's pole.
This pole correction is relevant only for the piece which looks like
a Lippmann-Schwinger
equation iterate and, hence, we will drop the recoil correction everywhere except for this
one contribution.  These recoilless
propagators depend only on the energy and are
given by $(p_0 + i \epsilon)^{-1}$.

A tedious calculation yields
\begin{eqnarray}
i {\cal A}_{a,b,c} & = &g_{1m}^4 g_{2m}^{}
\int
\prod_{j = 1}^{3} \frac{d^3 \mbox{\boldmath$k$}_j}{(2 \pi)^3}
\, (2 \pi)^3 \, \delta^{(3)}
\left( \sum_{j = 1}^{3}
\mbox{\boldmath$k$}_j - \mbox{\boldmath$q$} \right)
\times\!\!\
 \int \prod_{j = 1}^{3} \frac{d \omega_j}{2 \pi} \,
(2 \pi) \delta \left( \sum_{j = 1}^{3} \omega_j  \right ) \nonumber \\
& \times & \,  D(k_1) \,
\left [  \delta \left( p_0 - \omega_1 + (\mbox{\boldmath$p$}+
 \mbox{\boldmath$k_1$})^2 /(2 M_N) \right ) \,
G(\tilde{p} + k_1) \right ] \, \nonumber \\
& \times & \, D(k_2)D(k_3) \, \frac{1}{\omega_1 + \omega_2
+ i \epsilon} + {\cal O}(N_c) \, ,
\label{result}
\end{eqnarray}
where the ${\cal O}(N_c)$ corrections come from using nonrelativistic propagators and neglecting recoil.
The combination in square brackets can easily be seen to be the delta function fixing $\omega_1$
times the Lippmann-Schwinger propagator,
$G_{LS}(\mbox{\boldmath$p$}, E)  = \frac{1}{E - \mbox{\boldmath$p$}^2/M_N + i \epsilon}$.

Note that the delta function restricts $\omega_1$ to being of order $N_c^{-1}$ for these
kinematics.  The expression in square brackets is sensitive to the infrared kinematics and
depends on the fact that $\omega_1 \ne 0$.
However, for all parts of the expression, setting
$\omega_1$ to zero will only induce an error of relative size $(1/N_c)$.   Substituting zero for
$\omega_1$ in all parts of eq.~(\ref{result}) except for the factor in square brackets gives
\begin{eqnarray}
i {\cal A}_{a,b,c} & = &\int \frac{d^3 \mbox{\boldmath$k$}_1 }{(2 \pi)^3} \,
V_1(\mbox{\boldmath$k$}_1) \, G_{LS}(\mbox{\boldmath$k$}_1, E) \, V_2(\mbox{\boldmath$q$} -
\mbox{\boldmath$k$}_1 ) + { \cal O}(N_c) \nonumber \\
 V_1(\mbox{\boldmath$k$}) \,& = &\,
\frac{g_{1m}^2}{ \mbox{\boldmath$k$}^2 + m_m^2} \nonumber \\
V_2 (\mbox{\boldmath$k$})  \, &= &\,
g_{1m}^2 g_{2m} \int \prod_{j = 1}^{3} \frac{d^3 \mbox{\boldmath$k$}_j}{(2 \pi)^3} \, \int \,
\prod_{j=2}^3 \frac{d \omega_j}{2 \pi} \, (2 \pi)^3 \, \delta^{(3)}\left( \sum_{j = 2}^{3}
\mbox{\boldmath$k$}_j - \mbox{\boldmath$k$} \right) (2 \pi) \delta \left( \sum_{j = 2}^{3}
\omega_j  \right ) \nonumber\\
& \times & D(k_2)D(k_3) \frac{1}{\omega_2 + i \epsilon} \; \; .
\label{potform}
\end{eqnarray}
Note that $V_1$ is a static one-meson exchange potential
while $V_2$ is the static
potential for the exchange of two mesons with one seagull vertex.  Thus, the form of
eq.~(\ref{potform}) is precisely part of an iterate of the
Lippmann-Schwinger equation, with $V_1$ as the first iteration
and  $V_2$ the second.  The other ordering comes from the graphs
(e), (f) and (g).  Since the part of $
i {\cal A}_{a,b,c}$ which is of order $N_c^2$ is a static potential iterate, one concludes the
contributions of these graphs to the potential is necessarily
of order $N_c$ or less and, hence, is
consistent with the $N_c$ scaling rules of eq.~(\ref{KSM}).

Note that had one followed the organization of ref.~\cite{BelCoh02} there would have been a
remaining contribution to the potential of order $N_c^2$.  From this we see an explicit example
where the energy-dependent potentials extracted from a set of Feynman graphs associated with
multiple-meson exchange using the algorithm of ref.~\cite{BelCoh02} are inconsistent with the
large $N_c$ counting rules of refs.~\cite{KapSav95,KapMan96} while energy-independent ones are
consistent. Thus, we see the large $N_c$ nuclear potential puzzle has been resolved for this set
of diagrams. Although we have not explicitly calculated other multiple-meson exchange graphs such
as the sum of ladders and crossed-ladders considered in ref.~\cite{BelCoh02} due to their
complexity, it is highly plausible that consistency would be obtained for the energy-dependent
potentials for such cases.

In summary, it has been shown that energy-dependent potentials  can have different $N_c$ scaling
behavior than energy-independent potentials.  It has been argued that the $N_c$ scaling rules of
eq.~(\ref{KSM}) should apply only to
energy-independent potentials.  With these energy-dependent
potentials one expects consistency.   From this perspective, the
inconsistency of the three- and higher-meson exchange
potentials in ref.~\cite{BelCoh02} with
eq.~(\ref{KSM}) can be understood as arising from the fact that the analysis used in that work
yielded energy-dependent potentials.  This interpretation is highly plausible given the explicit
demonstration of consistency for the case of the energy-independent potential associated with
 three-scalar-meson exchange of the type seen in fig.~(\ref{seagull}) despite the fact that the
energy-dependent potential was obtained  using the methods of ref.~\cite{BelCoh02}

This work is supported by the U.S.~Department of Energy grant
DE-FG02-93ER-40762. The author gratefully acknowledges useful discussions with
 A. Beltisky, B. Gelman, D. Kaplan, A. Manohar and D. Riska.


\begin{thebibliography}{100}

\bibitem{Hoo74}G. 't Hooft,
Nucl. Phys. B 72 (1974) 461.
\bibitem{Wit79}
E. Witten,
Nucl. Phys. B 160 (1979) 57.
\bibitem{BanCohGel01}
M. K. Banerjee, T. D. Cohen, B. A. Gelman,
Phys. Rev. C {\bf 65} 034011 (2002).
\bibitem{KapMan96}
D.B. Kaplan, A.V. Manohar,
Phys. Rev. C {\bf 56} (1997) 76.
\bibitem{KapSav95}
D.B. Kaplan, M.J. Savage,
Phys. Lett. {\bf B 365} (1996) 244.
\bibitem{BelCoh02}
A. V. Belitsky and  T. D. Cohen  hep-ph/0202153.
\bibitem{CohGel02}
T. D. Cohen, B. A.  Gelman, nucl-th/0202036
\bibitem{GerSak83}
J.L. Gervais, B. Sakita,
Phys. Rev. Lett. {\bf 52} (1984) 87.
\bibitem{GerSak84}
J.L. Gervais, B. Sakita,
Phys. Rev. D {\bf 30} (1984) 1795.
\bibitem{DasMan93}
R.F. Dashen, A.V. Manohar,
Phys. Lett. {\bf B 315} (1993) 425.
\bibitem{DasJenMan93}
R. F. Dashen, E. Jenkins, A.V. Manohar,
Phys. Rev. D {\bf 49} (1994) 4713, (E) D {\bf 51} (1994) 2489.
\bibitem{Jen93}
E. Jenkins,
Phys. Lett. {\bf B 315} (1993) 441.
\bibitem{Ris02}
D. O. Riska nucl-th/0204016.
\bibitem{Bea02}
S. Beane, hep-ph/0204107.
\bibitem{Coh89} Thomas D. Cohen, Phys. Rev. Lett. {\bf 62}
(1989) 3027.
\bibitem{LamLiu96}
C.S. Lam, K.F. Liu,
Nucl. Phys. B 483 (1997) 514.
\bibitem{LamLiu97}
C.S. Lam, K.F. Liu,
Phys. Rev. Lett. 79 (1997) 597.
\bibitem{ManStew} A. V. Manohar and I. W. Stewart, Phys. Rev. D 62 (2000) 074015;
 Phys. Rev. D63 (2001) 054004.

\end{thebibliography}
\end{document}